\documentclass[conference]{IEEEtran}

\IEEEoverridecommandlockouts

\usepackage{amsmath,amssymb,amsfonts}
\usepackage{algorithmic}
\usepackage{graphicx}
\usepackage{xcolor}
\usepackage{multirow}
\usepackage{float}
\usepackage{xspace}
\usepackage[capitalize]{cleveref}
\usepackage{makecell}
\usepackage{bbding}
\usepackage{url}
\usepackage{booktabs}
\usepackage{pdfpages}
\usepackage{cite}
\usepackage{fancyhdr}

\renewcommand\_{\textunderscore\allowbreak}

\newcommand{\nickname}[1]{AlphaSparse} 

\newcommand{\inds}[1]{\ensuremath{\mathrm{#1}}\xspace}

\newcommand{\changed}[1]{{#1}}

\def\BibTeX{{\rm B\kern-.05em{\sc i\kern-.025em b}\kern-.08em
    T\kern-.1667em\lower.7ex\hbox{E}\kern-.125emX}}
\begin{document}

\title{AlphaSparse: Generating High Performance SpMV Codes Directly from Sparse Matrices}


\author{\IEEEauthorblockN{Zhen Du\IEEEauthorrefmark{1}\IEEEauthorrefmark{3}, Jiajia Li\IEEEauthorrefmark{2}, Yinshan Wang\IEEEauthorrefmark{1}, Xueqi Li\IEEEauthorrefmark{1}, Guangming Tan\IEEEauthorrefmark{1}, Ninghui Sun\IEEEauthorrefmark{1}}
\IEEEauthorblockA{
\IEEEauthorrefmark{1} \textit{Institute of Computing Technology, Chinese Academy of Sciences, Beijing, China}}
\IEEEauthorblockA{
\IEEEauthorrefmark{2} \textit{North Carolina State University, Raleigh, NC, USA}}
\IEEEauthorblockA{
\IEEEauthorrefmark{3} \textit{University of Chinese Academy of Sciences, Beijing, China}}

\IEEEauthorblockA{\{duzhen18z,wys,lxq,tgm,snh\}@ict.ac.cn, jli256@ncsu.edu}
}


\maketitle

\thispagestyle{fancy}
\lhead{}
\rhead{}
\chead{}
\lfoot{
\footnotesize{
SC22, November 13-18, 2022, Dallas, Texas, USA
\newline 
978-1-6654-5444-5/22/\$31.00 \copyright 2022 IEEE}}
\rfoot{}
\cfoot{}
\renewcommand{\headrulewidth}{0pt} \renewcommand{\footrulewidth}{0pt}

\begin{abstract}

Sparse Matrix-Vector multiplication (SpMV) is an essential computational kernel in many application scenarios. Tens of sparse matrix formats and implementations have been proposed to compress the memory storage and speed up SpMV performance. We develop AlphaSparse, a superset of all existing works that goes beyond the scope of human-designed format(s) and implementation(s). AlphaSparse automatically \emph{creates novel machine-designed formats and SpMV kernel implementations} entirely from the knowledge of input sparsity patterns and hardware architectures. Based on our proposed Operator Graph that expresses the path of SpMV format and kernel design, AlphaSparse consists of three main components: Designer, Format \& Kernel Generator, and Search Engine. It takes an arbitrary sparse matrix as input while outputs the performant machine-designed format and SpMV implementation. By extensively evaluating 843 matrices from SuiteSparse Matrix Collection, 
AlphaSparse achieves significant performance improvement by 3.2$\times$ on average compared to five state-of-the-art artificial formats and 1.5$\times$ on average (up to 2.7$\times$) over the up-to-date implementation of traditional auto-tuning philosophy.

\end{abstract}


\section{Introduction}\label{introduction}


Sparse Matrix-Vector multiplication (SpMV, \emph{y=Ax}) is one of the most computational kernels in many domains, such as climate simulation~\cite{tillenius2013task}, computer graphics~\cite{weber2013efficient}, molecular dynamics\cite{zheng2013algorithms, kylasa2014puremd}, data analytic~\cite{chen2018gflink, kyrola2012graphchi}, machine/deep learning~\cite{wang2019deep, zhang2016cambricon}, to name a few. In the past decades, many efforts have been conducted to improve SpMV performance through proposing sparse matrix formats, leveraging various performance optimization methods, and automatic performance tuning (auto-tuning).

\changed{Dozens of sparse matrix formats have been proposed to efficiently compress sparse matrices in memory on contemporary architectures: multi-core CPUs~\cite{660313}, Graphics Processing Units (GPUs)~\cite{owens2008gpu}, Intel Xeon Phi accelerators~\cite{sodani2016knights}, and Field-Programmable Gate Array (FPGAs)~\cite{kuon2008fpga}. These formats are designed for diverse goals: reducing memory access, improving load balance, reducing GPU thread divergence, etc. They store only non-zero elements and ignore zeros which take a major portion of a sparse matrix. (Refer to the work~\cite{langr2015evaluation} and~\cite{filippone2017sparse} for good summaries of them). We categorize sparse matrix formats into three groups: Root Formats, Derived Formats, and Hybrid Formats.


Four formats are generally considered as basic formats~\cite{li2013smat, filippone2017sparse, xie2019ia}, or \emph{Root Formats}, which consists of \texttt{COOrdinate (COO)}, \texttt{Compressed Sparse Row (CSR)}, \texttt{ELLPACK (ELL)}, and \texttt{DIAgonial (DIA)}. To handle more irregular matrices, better memory compression, or higher performance of sparse kernels, plenty of \emph{Derived Formats} have been proposed. We refer to a derived format as a format manually designed based on only ONE root format, such as \texttt{Blocked COORdinate (BCOO)}~\cite{im2004sparsity}, derived from \texttt{COO}; \texttt{CSR5}~\cite{liu2015csr5}, derived from \texttt{CSR}; \texttt{Sliced ELLPACK (SELL)}, derived from \texttt{ELL}, to name a few. Beyond root and derived formats, \emph{Hybrid Formats} flexibly use multiple formats for different portions of a sparse matrix. It could be a mix among root and derived, such as \texttt{HYBrid (HYB)}, \texttt{COCKTAIL}~\cite{su2012clspmv}, and \texttt{Compressed Sparse eXtended (CSX)}~\cite{kourtis2011csx}.}

Because of the diversity of sparsity patterns and close association between input matrix features, architecture characteristics, and SpMV performance, it is unrealistic to find a one-fits-all format or optimization method. 
Thus, SpMV auto-tuners such as SMAT~\cite{li2013smat}, clSpMV~\cite{su2012clspmv}, Zhao et al.~\cite{zhao2018bridging}, have been designed to select the most appropriate format for a given matrix from a set of candidate artificial formats.

Despite the efforts of all researches mentioned above, this classic but stubborn kernel is still largely behind its attainable performance from analysis, especially for highly irregular sparse data~\cite{daga2015structural}. We observe three problems in state-of-the-art researches preventing SpMV from achieving higher performance.


\textbf{Problem 1: Limited human practices meet an ever-growing number of sparse matrices.} Generally, a matrix format could handle only a specific matrix pattern and perform this type of matrix well. Thus, the patterns not covered in this format lead to low performance. According to our experiments in SuiteSparse Matrix Collection~\cite{10.1145/2049662.2049663}, there is an approximate $10\times$ maximum-minimum performance gap observed from mainstream formats \texttt{ELL}, \texttt{HYB}, \texttt{ACSR}~\cite{ashari2014fast} and \texttt{CSR-Adaptive}~\cite{daga2015structural}. From another point of view, SuiteSparse Matrix Collection has gradually collected 2893 matrices from 91 domains. As domains and data emerge from real-world problems, most probably, we will face unseen sparse data and patterns in the future, which would need new formats and kernel implementations. It is not effective, practical, or even possible for researchers to keep designing new formats for any incoming matrices.


\textbf{Problem 2: Challenge of irregular sparsity.} Irregularity is almost the biggest challenge in nowadays SpMV program design. It brings a great diversity of distributions for row lengths\changed{\footnote{\changed{For sparse matrix, row length is the number of non-zeros in a row.}}} and row positions, which causes enormous difficulties for efficient parallelism and memory access~\cite{daga2015structural}. In this paper, we define sparse matrices where the variances of its row lengths are more than 100 as irregular matrices, according to target matrices of recent format studies~\cite{yan2014yaspmv, 10.1145/3437801.3441592}. Irregular matrices occupy more than 35\% of SuiteSparse Matrix Collection. General sparse matrix formats cannot accommodate irregular sparsity well due to highly redundant computing, unbalanced load, memory access hot-spots, etc. Though some new formats~\cite{liu2015csr5, merrill2016merge, ashari2014fast} have been proposed, particularly for irregular sparsity, they still have limited applicability (they only focus on 10-20 matrices in their evaluation).




\begin{figure}[htbp]
\vspace{-0.5em}
\centering
\scriptsize
\begin{tabular}{c}
\includegraphics[width=0.38\textwidth]{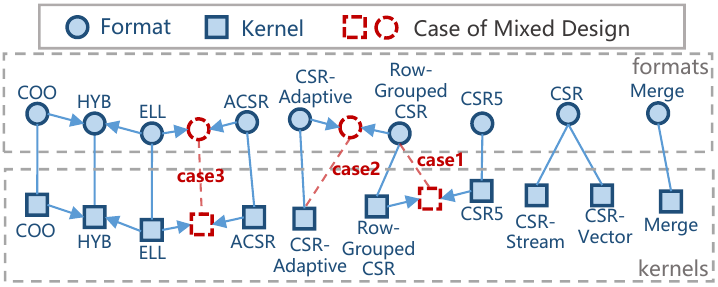}\\
(a)\\
\includegraphics[width=0.35\textwidth]{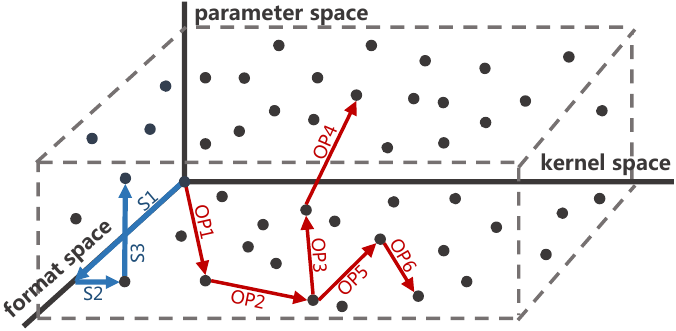}\\
(b)\\
\end{tabular}
\vspace{-0.5em}
\caption{(a) Search space of a traditional auto-tuner. (b) Searching methodologies of format-selection auto-tuner and AlphaSparse in origin design space.}
\label{DesignSpace}
\vspace{-0.5em}
\end{figure}

\textbf{Problem 3: Limitation of existing auto-tuners.} 
Existing auto-tuners are modeled as coarse-grained format selectors~\cite{li2013smat,su2012clspmv} and are limited by human experience and implementations. We depict a small set of artificial formats and SpMV implementations (blue circles and squares, respectively) in \Cref{DesignSpace}a. Each format has one or more coupled kernel implementations. Traditional auto-tuners essentially choose a format-kernel combination (represented as solid blue lines), where other potentially existing but humanly undiscovered formats, kernel implementations, and the connections in-between (shown in cases 1-3 in red) have been overlooked. Take case 1 as an example, it uses the existing \texttt{row-grouped CSR} format \cite{oberhuber2010new} but with a new implementation by combining thread-level reduction of \texttt{CSR5} \cite{liu2015csr5} and global memory reduction of \texttt{row-grouped CSR}. These omitted cases cause an auto-tuner to miss opportunities for potential performance improvement. To make things worse, complexity from irregular sparsity amplifies this shortcoming of format selectors.


AlphaSparse solves these problems by targeting one ultimate goal: \textbf{creating machine-designed SpMV programs that surpass the scope of human practices and outperform both artificial formats and traditional auto-tuners.} We achieve this goal by directly searching in the original design space of the SpMV program, which contains three dimensions: 1) format, the data layout in memory; 2) kernel, the way that data is calculated; 3) parameter, the quantitative details of the first two dimensions (illustrated in \Cref{DesignSpace}b). Every position of the design space represents an SpMV program. The blue path shows the selection strategy of traditional auto-tuners that can only take steps in parallel with any of the three directions. In contrast, AlphaSparse proposes a new model, named \emph{Operator Graph}, which simulates the SpMV design philosophy to exploit much larger space. An Operator Graph is a ``path" to a specific location of the design space by connecting arbitrary numbers of operators (detailed in \Cref{DesignSpaceExpression}). An operator, a vector in design space, represents a design strategy of the SpMV program and can simultaneously ``move" in three dimensions. This more flexible and integrated model enables AlphaSparse to reach designs inaccessible to existing human works and gain more opportunities for higher performance. 

\begin{table}[htbp]
\vspace{-0.5em}
\caption{Comparison of AlphaSparse to state-of-the-art works.}
\vspace{-0.5em}
\scriptsize
\centering
\begin{tabular}{ccccc}
\toprule
\multicolumn{2}{c}{Work} & \textbf{Sparsity} & \textbf{Irregularity} & \textbf{Creativity}\changed{\footnotemark}  \\
\midrule
\multirow{2}*{\makecell[c]{Artificial \\ Format \\ Designs}} & \makecell[c]{CSR,\\ELL,\\COO,etc.} & \Checkmark  & \XSolidBrush & \XSolidBrush \\
\cline{2-5}
& \makecell[c]{\makecell{CSR5~\cite{liu2015csr5},\\Merge~\cite{merrill2016merge},\\ACSR~\cite{ashari2014fast},etc.}} & \Checkmark  & \Checkmark & \XSolidBrush \\
\midrule
\makecell[c]{Traditional \\ Auto-tuners} & \makecell[c]{SMAT~\cite{li2013smat},\\clSpMV~\cite{su2012clspmv},\\Zhao et al.~\cite{zhao2018bridging}} & \Checkmark  & \XSolidBrush \footnotemark & \XSolidBrush \\
\midrule
\multirow{2}*{\makecell[c]{Compiler \\ Technologies}} & \makecell[c]{TVM~\cite{chen2018tvm}} & \XSolidBrush  & \XSolidBrush & \XSolidBrush \\
& \makecell[c]{TACO~\cite{kjolstad2017tensor}} & \Checkmark  & \XSolidBrush & \XSolidBrush \\
\midrule
\midrule
\makecell[c]{Intelligent \\ Auto-tuner} & \makecell[c]{AlphaSparse} & \Checkmark  & \Checkmark & \Checkmark \\
\bottomrule
\end{tabular}
\label{ComparationSummary}
\vspace{-0.5em}
\end{table}

\footnotetext[2]{\changed{The ability to create new machine-designed SpMV formats and kernels.}}

\Cref{ComparationSummary} compares AlphaSparse with mainstream related works from angles of sparsity, irregularity, and creativity. Compared with artificial format designs and traditional SpMV auto-tuners, AlphaSparse shows its novelty in creativity and irregularity. It is the first work that creates completely novel machine-designed formats along with their SpMV implementations to pursue high performance. Some compilers seem to be more flexible, especially TACO~\cite{kjolstad2017tensor}. However, its general IR (intermediate representation) hides details of algorithms and hardware architectures, which covers only basic optimizations for general sparse problems and misses many optimization opportunities. Besides, TACO still explores limited artificial formats by leveraging the ``level formats" concept for each dimension~\cite{chou:2018:formats,chou:sm-thesis:2018}, same as format selectors.


\footnotetext{Zhao et al. partly solves the irregularity by including \texttt{CSR5} format.}

However, three challenges need to be conquered to build the intelligent AlphaSparse. The first one is a much larger search space. Let $\mathcal{A}$ be the number of all known artificial formats and assume each of them provides a unique format or kernel design strategy. By only comparing the format-kernel subset of search space, its size of traditional auto-tuning is $O(\mathcal{A})$, while $O(\mathcal{A}^p)$ theoretically in AlphaSparse with an Operator Graph including $p$ Operators. The second is integrated modeling. Extracting design strategies of SpMV from a large number of existing works and expressing them in a unified IR is non-trivial. The last challenge is projecting positions in the origin design space to three dimensions to obtain corresponding SpMV programs.

AlphaSparse has three main components to solve these challenges: Designer, Format \& Kernel Generator, and Search Engine, to accomplish design space’s expression, projection, and exploration. Designer and Format \& Kernel Generator accept Operator Graphs as input and generate formats with corresponding kernel implementations. Search Engine aims at finding an Operator Graph with high performance. While searching, SpMV performance corresponding to Operator Graph can be obtained by directly running the generated SpMV program.
We implement AlphaSparse in more than 110,000 lines of C++ codes that will be released. Although we only focus on SpMV in this paper, the methodology of AlphaSparse can even adapt to more sparse problems by defining new corresponding operators and backends.

Our main contributions are summarized as follows: 


\begin{itemize}
    \item We first show potential high-performance SpMV programs overlooked in existing works and the necessity and feasibility foundations for AlphaSparse (\Cref{Background}). 
    \item We develop AlphaSparse, which is easy to use by taking Matrix Market files as input and outputting high-performance SpMV codes generated by the machine.
    AlphaSparse can be considered as a counterpart of \emph{AlphaFold}~\cite{jumper2021highly}, which predicts the protein structure from the beginning, in high-performance sparse problems; while traditional auto-tuners correspond to traditional template-based methods in protein structure prediction. (\Cref{Overview})
    \item The design space is expressed by a newly proposed graph-based modeling, called Operator Graph (\Cref{DesignSpaceExpression}); projected by format and kernel generators to generate compressed data representation and high-performance implementation (\Cref{DesignSpaceProjection}); and explored by a three-level search and pruning strategies (\Cref{DesignSpaceExploration}) in AlphaSparse.
    \item We evaluate AlphaSparse on 843 large matrices from SuiteSparse Matrix Collection. AlphaSparse largely improves SpMV performance by up to 22.2$\times$ (3.2$\times$ on average) compared to five human-designed state-of-the-art formats. We also compare AlphaSparse with an up-to-date implementation of format selector, where AlphaSparse achieves up to 2.7$\times$ (1.5 $\times$ on average) performance improvement. (\Cref{evaluation})
\end{itemize}

\section{Motivation} \label{Background}

The motivation of AlphaSparse comes from two observations, which separately show its necessity and feasibility.

\begin{figure}[htbp]
\vspace{-0.5em}
\centering
\includegraphics[width=0.46\textwidth]{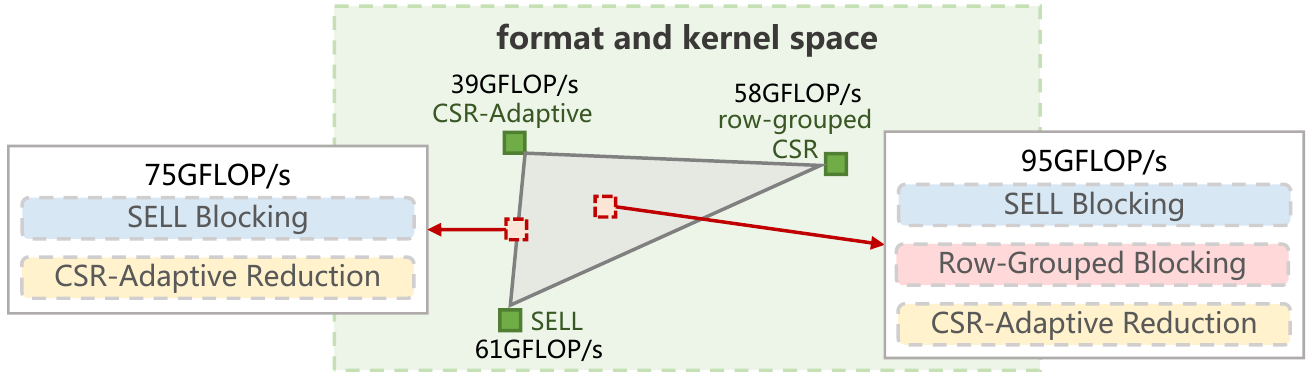}
\vspace{-1em}
\caption{Mixed designs found by AlphaSparse on the matrix \inds{2D\_27628\_bjtcai} in the space of format and kernel.}
\label{MixedDesignPerformanceCase}
\vspace{-1em}
\end{figure}

\emph{Observation 1: Artificial formats and their sparse kernel algorithms are limited by human experience and narrow search space, which misses the potential for higher performance.} Newly proposed artificial sparse matrix formats and auto-tuners have covered increasing sparse patterns. However, human practice ignores a large number of potential formats and kernels. As shown in \Cref{MixedDesignPerformanceCase}, on matrix \inds{2D\_27628\_bjtcai} from SuiteSparse Matrix Collection, \texttt{CSR-Adaptive}~\cite{greathouse2014efficient}, \texttt{row-grouped CSR}~\cite{heller2012adaptive}, \texttt{SELL}~\cite{kreutzer2012sparse} separately achieves 39 GFLOPS, 58 GFLOPS and 61 GFLOPS. By combining the blocking strategy of \texttt{row-grouped CSR} with the reduction strategy of \texttt{CSR-Adaptive}, the performance of the mixed format is higher as 75 GFLOPS. Similarly, by mixing formats and kernels from all these source formats, the performance could be even higher as 95 GFLOPS.


\begin{figure}[htbp]
\vspace{-0.5em}
\centering
\includegraphics[width=0.35\textwidth]{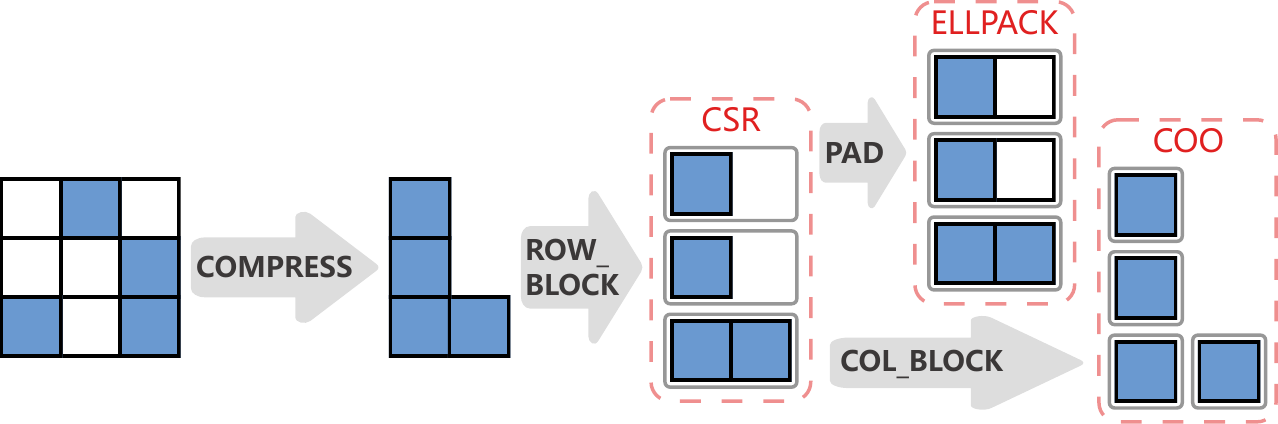}
\vspace{-1em}
\caption{The steps of converting a tiny sparse matrix to \texttt{CSR}, \texttt{COO} and \texttt{ELL}. Blue blocks are non-zeros, while blank ones are zeros.}
\label{GraphofCSRCOOELL}
\vspace{-0.5em}
\end{figure}

\begin{figure*}[htbp]
\vspace{-0.5em}
\centering
\includegraphics[width=0.8\textwidth]{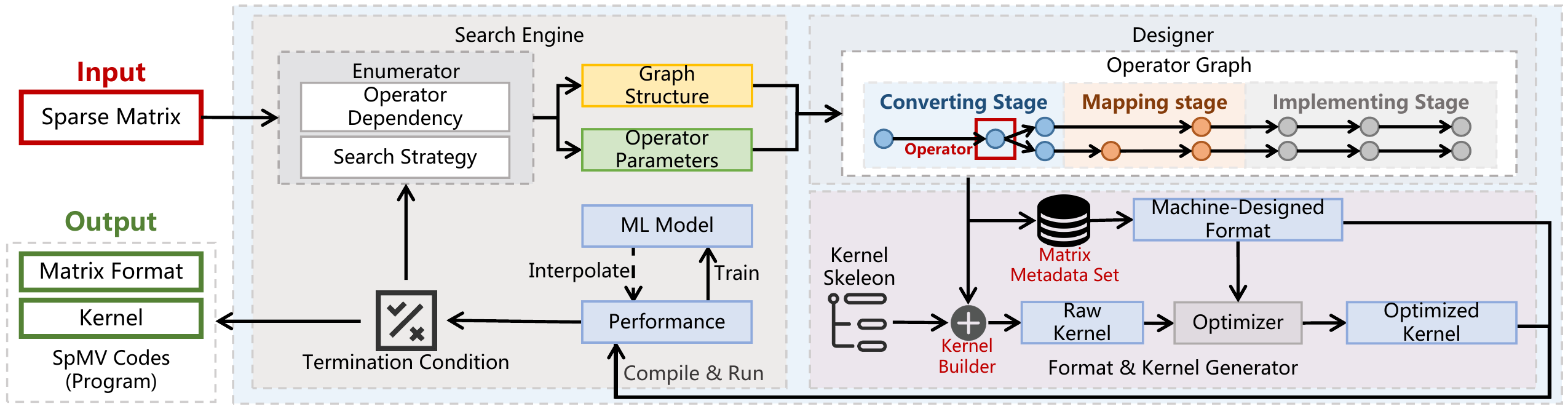} 
\vspace{-1em}
\caption{Overview of AlphaSparse.}
\label{AlphaSparseOverview} 
\vspace{-1em}
\end{figure*}

\emph{Observation 2: Sparse formats are converted from the source matrix with common steps, making creating new formats feasible from more combinations of these common steps.} This observation has been proved by other work~\cite{chou2020automatic}, although they underscored the conversion among existing artificial formats.
Usually, when a new artificial format is designed, the conversion routine will also be provided from the original matrix. We take the conversion of three root formats as examples, shown in \Cref{GraphofCSRCOOELL}. In the beginning, the original input matrix is compressed by ignoring all zeros. By blocking the matrix in each row, \texttt{CSR} format can be obtained. Furthermore, by further padding in each block or by blocking in each column, \texttt{ELL} or \texttt{COO} can be generated. These four steps commonly exist in other format conversions~\cite{anzt2014implementing, heller2012adaptive, kreutzer2014unified}. Thus, it is feasible to generate or even create a format automatically by taking more common conversion steps.


\section{Overview} \label{Overview}




AlphaSparse proposes an integrated model named Operator Graph. Operator Graph describes and explores the origin three-dimensional design space of format, kernel and parameter simultaneously with operators (shown in \Cref{DesignSpace}). It provides a meticulous search to handle the complexity brought by sparsity patterns that are highly associated with SpMV performance. \textbf{An operator uniformly expresses the information of kernel and format design, including the configurations of their parameters.} Through transforming Operator Graphs, not only high-performance but also new machine-designed formats and kernels could be generated.

AlphaSparse consists of a \emph{Search Engine} (\Cref{DesignSpaceExploration}), a \emph{Designer} (\Cref{DesignSpaceExpression}),  and a \emph{Format \& Kernel Generator} (\Cref{DesignSpaceProjection}). As shown in \Cref{AlphaSparseOverview}, the Search Engine first enumerates Operator Graphs by generating graph structures and corresponding parameters for their operators under a given search strategy. The enumerated Operator Graph will be sent to the Designer. The Designer executes these operators in order to modify the Matrix Metadata Set, which includes all details of the matrix state (detailed in \Cref{MatrixMetadataSet}). At last, Format \& Kernel Generator produces the kernel and format according to the Operator Graph and Matrix Metadata Set, with several optimizations (detailed in \Cref{DesignSpaceProjection}). For a specific structure of Operator Graph, AlphaSparse first gets its performances by directly running the SpMV program of each parameter combination on a coarse-grained grid. To further achieve a detailed search in parameter space with low overhead, AlphaSparse uses a lightweight machine learning (ML) cost model to interpolate parameters to a fine-grained grid. Till the termination condition based on simulated annealing is satisfied, the search process stops and outputs the best SpMV codes found by it.


AlphaSparse has already provided high out-of-the-box performance and is easy to use for top-level users. Users only need to input a Matrix Market file of a sparse matrix, and AlphaSparse will output a matrix stored in a specific format and a kernel implementation. Essentially, apart from traditional auto-tuners, AlphaSparse is moving forward a significant step by acting as a substitute for algorithm researchers in developing new SpMV formats and kernels. Usually, this kind of algorithm work not only highly depends on individual inspiration but also costs time of either months or years. AlphaSparse only takes hours to greatly outperform almost all artificial designs. From this aspect, AlphaSparse is not a traditional online performance tuner but a tool for the SpMV algorithm research or an extremely optimized library generator, which narrows the focus from the entire algorithm to a particular operator(s). The generated codes can be directly called in real-world applications. The artifact description of this paper shows its usage.



\begin{table*}
\vspace{-0.5em}
\caption{Operators considered in \nickname{}.}
\centering
\scriptsize
\resizebox{0.91\textwidth}{!}{
\begin{tabular}{cllp{.53\textwidth}}
\hline
\textbf{Stage} & \textbf{Operator} & \textbf{Source} & \textbf{Description} \\
\hline
\multirow{5}*{\textbf{Converting}} & ROW(COL)\_DIV & \cite{10.1145/2464996.2465013,7875496} & Divide a matrix in rows/columns \\
& SORT & \cite{cao2010implementing, kreutzer2012sparse} & Sort rows in decreasing order of \#non-zeros per row \\
& SORT\_SUB & \cite{zheng2014biell, cao2010implementing, kreutzer2012sparse} & Sort rows in decreasing order of \#non-zeros per row with in a submatrix\\
& BIN & \cite{ashari2014fast, hou2017auto} & Put rows into different bins according to \#non-zeros per row \\
& COMPRESS & \cite{naumov2010cusparse} & Ignore all zeros of the sparse matrix \\
\hline
\multirow{4}*{\textbf{Mapping}} & BMTB(BMW,BMT)\_ROW(COL)\_BLOCK & \cite{ashari2014efficient, kreutzer2014unified,zheng2014biell, maggioni2013adell} & Split a matrix in row/column dimension, each of which mapped to a thread block/warp/thread \\
& BMT\_NNZ\_BLOCK & \cite{yan2014yaspmv, liu2015csr5, 7875496} & Map continuous non-zeros to threads \\
& BMTB(BMW,BMT)\_PAD & \cite{maggioni2013adell, heller2012adaptive, ashari2014efficient} &  Zero padding to BMTB/BMW/BMT \\
& SORT\_BMTB & \cite{kreutzer2014unified} & Sorting rows in decreasing order of \#non-zeros per row within a BMTB\\
\hline
\multirow{9}*{\textbf{Implementing}} & SET\_RESOURCES & / & Set runtime configurations \\
& GMEM\_ATOM\_RED & \cite{heller2012adaptive} & Atomically add intermediate results to global memory \\
& SHMEM\_OFFSET\_RED & \cite{daga2015structural, greathouse2014efficient, merrill2016merge} & Reduce intermediate results from multiple rows to shared memory, according to row offset \\
& SHMEM\_TOTAL\_RED & \cite{ashari2014fast, daga2015structural} & Reduce intermediate results of the same row in shared memory \\
& WARP\_TOTAL\_RED & \cite{yang2012improved, liu2015lightspmv} & Reduce all the intermediate results per warp to one row \\
& WARP\_BITMAP\_RED & \cite{maggioni2013adell} & Reduce all the intermediate results per warp by bitmap \\
& WARP\_SEG\_RED & \cite{liu2015csr5} & Reduce all the intermediate results per warp by segment sum \\
& THREAD\_TOTAL\_RED & \cite{ashari2014fast, maggioni2013adell, feng2011optimization} & Reduce all the intermediate results per thread to one row \\
& THREAD\_BITMAP\_RED & \cite{yan2014yaspmv, liu2015csr5} & Reduce intermediate results per thread by bitmap \\
\hline
\multicolumn{4}{l}{BMTB/BMW/BMT is abbreviation of ``a block mapped to a thread block" or ``warp" or ``thread".}
\end{tabular}
}
\label{TheExampleofFormatDesignNode}
\vspace{-0.5em}
\end{table*}

\begin{figure*}[htbp] 
\vspace{-0.5em}
\centering
\includegraphics[width=0.9\textwidth]{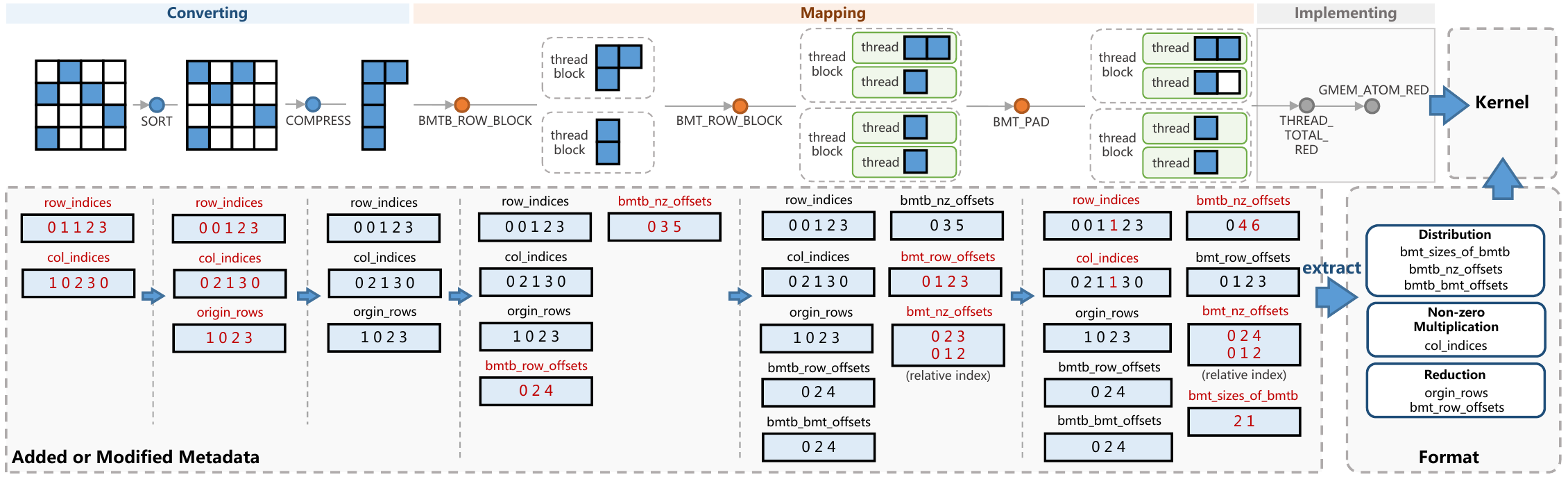} 
\vspace{-1em}
\caption{An Example of format generation. The upper is an Operator Graph; the lower is a subset of Matrix Metadata Set.}
\label{ExampleofFormatGeneration} 
\vspace{-1em}
\end{figure*}

\section{Designer} \label{DesignSpaceExpression}


The \emph{Designer} maintains the \emph{Operator Graph}, the key data structure of AlphaSparse. We are the first to break existing formats and kernel implementations~\cite{langr2015evaluation, filippone2017sparse} into finer-grained design strategies and use them to model the SpMV program (shown in \Cref{TheExampleofFormatDesignNode}). As the combination of operators, Operator Graph opens a wider integrated space of format and kernel designs. Compared to existing format selectors, AlphaSparse possesses higher flexibility for performance tuning, thus obtaining outperforming SpMV codes in larger probabilities.



\subsection{Operator} \label{Operator}

Given a sparse matrix, we summarize that its SpMV program is generally developed in three steps: 1) defining a compressed memory layout (i.e., format) of the matrix; 2) mapping (distributing) it to hardware units of different parallelism levels; 3) designing kernel implementation, mainly SpMV reduction strategies. These stages are \emph{converting}, \emph{mapping} and \emph{implementing}. Each stage consists of multiple design or optimization strategies, called \emph{operators}\footnote{Operators in AlphaSparse represent designs of format and kernel implementation, different from mathematical operators.}. Defining operators is non-trivial and challenging, which needs plenty of preparatory work to abstract optimizing strategies from existing works and validate their effectiveness in the final performance. For prototyping purposes, AlphaSparse currently only considers operators for GPUs. We list all the operators in AlphaSparse in \Cref{TheExampleofFormatDesignNode}. Almost all of them are derived from existing research, as shown in the ``Source" column. \changed{At the level of the overall design of the SpMV program, AlphaSparse has covered the whole design process by the three stages of Operator Graph. At the level of design strategies (so-called operators), it is not easy to get their quantitative and theoretical coverage. As far as we know, AlphaSparse has covered almost all popular formats with high performances}.

Operators in the converting stage define compressed memory layout. ROW(COL)\_DIV divides the whole matrix into striped sub-matrices in a row or column direction, which branches in the Operator Graph. Each sub-matrix can be treated separately in the following designs (shown on the upper right of \Cref{AlphaSparseOverview}) that help handle highly irregular matrices. SORT, SORT\_SUB, and BIN reorder matrix rows according to their lengths. COMPRESS ignores all zeros of a sparse matrix for storage.

The mapping stage always begins after the COMPRESS operator. Operators suffixed by \_BLOCK cut adjacent non-zeros of the matrix into blocks and map them to different levels of parallelism. The other operators in this stage further trim the memory layout inside of blocks. Operators suffixed by \_PAD add zeros to specific positions of a matrix to get more regular indices for higher performance. \changed{SORT\_BMTB reorders rows of each BMTB, which can reduce the range of sorting and create opportunities to decrease the padding rate.}.

Operators in the implementing stage are more relevant to kernel implementation. Except for SET\_RESOURCES, all the operators are suffixed by \_RED, which are different reduction strategies for intermediate results of BMTB, BMW, or BMT in an SpMV kernel. \changed{GMEM\_ATOM\_RED directly and atomically adds intermediate results to vector \emph{y} in global memory. Operators prefixed by SHMEM\_ are strategies for thread-block-level reduction in shared memory. SHMEM\_TOTAL\_RED fits for the condition where all intermediate results in a BMBT come from the same row. It adds up all intermediate results of a thread block to a result. SHMEM\_OFFSET\_RED includes CSR-like row offset indices~\cite{bell2008efficient} that record the position of the first intermediate result of each row in BMBTs. It reduces the intermediate results of each row in parallel.} Three operators prefixed by WARP\_ represent three mainstream strategies of warp-level reduction. WARP\_TOTAL\_RED is a classic strategy from CSR-Stream~\cite{daga2015structural}. For irregular matrices containing both short and long rows, WARP\_BITMAP\_RED and WARP\_SEG\_RED use \emph{bitmap}~\cite{maggioni2013adell} and \emph{segment sum}~\cite{blelloch1993segmented} to reduce results of BMW by rows. To gain more optimization opportunities from low-level details of the hardware, operators utilize a series of unique features of the GPU. In warp-level operators, hardware-level \emph{Warp Shuffle Functions}~\cite{CUDAToolkit} are used to achieve high performance of reduction. \changed{Operators prefixed by THREAD\_ are thread-level reductions in registers. THREAD\_TOTOAL\_RED is similar to other operators suffixed by \_TOTAL\_RED. THREAD\_BITMAP\_RED serially reduces the results of each row, using a bitmap to mark row boundaries.}

There is still a huge search space behind an operator that contains parameters of its details (parameter space in \Cref{DesignSpace}b), such as sorting granularity, the parallelism of reduction algorithms, blocking size, etc. Some design strategies derived from formats such as \texttt{HYB}, \texttt{CSB} \cite{bulucc2009parallel} are also critical to SpMV performance but have not been supported by AlphaSparse. AlphaSparse allows users to implement operators by themselves.

\subsection{Operator Graph} \label{OperatorGraph}

An Operator Graph is generated by connecting operators in order. The upper part of \Cref{ExampleofFormatGeneration} shows an elementary example. A real high-performance Operator Graph could be much deeper and sometimes include branches. This example mainly combines design philosophies of \texttt{SELL-P}~\cite{anzt2014implementing} and \texttt{CSR-Scalar}. COMPRESS, BMTB\_ROW\_BLOCK, BMT\_ROW\_BLOCK, BMT\_PAD, THREAD\_TOTAL\_RED, GMEM\_ATOM\_RED are from \texttt{SELL-P}, while COMPRESS, BMT\_ROW\_BLOCK, THREAD\_TOTAL\_RED, GMEM\_ATOM\_RED are from \texttt{CSR-Scalar}. SORT is from other formats, like JAD~\cite{li2013gpu}.




Dependencies exist between operators. They usually come from operators' semantics. Take the Operator Graph in \Cref{ExampleofFormatGeneration} as an example, BMT\_ROW\_BLOCK and BMT\_PAD cannot be followed by BMTB\_ROW\_BLOCK. Because when a data block has already been mapped to a thread, it cannot be further split and mapped to a thread block as higher-level parallelism in CUDA. Dependencies can also be defined by users for search pruning (detailed in \Cref{DesignSpaceExploration}).

\section{Format \& Kernel Generator} \label{DesignSpaceProjection}

Given an Operator Graph, we can move to a specific position of SpMV design space. To get the corresponding format and kernel implementation, Format \& Kernel Generator projects this position to format, kernel, and parameter space. Unlike traditional source code generators~\cite{chen2018tvm1}, which are based on a static template and only focus on the kernel implementation, Format \& Kernel Generator needs to handle flexible combinations of format and kernel by two components: \emph{Matrix Metadata Set} and \emph{Kernel Builder}.

\subsection{Matrix Metadata Set}\label{MatrixMetadataSet}

Matrix Metadata Set includes multi-perspective descriptions of the current matrix state, recording how the matrix is converted (detailed in observation 2 of \Cref{Background}). It is a huge key-value memory database whose contents are used to generate formats and kernels. Matrix Metadata Set contains basic matrix information (matrix size, number of columns and rows, length of each column and row), basic non-zero information (parent-block index, row index, column index), and information of blocks distributed to different levels of parallelism (block size, first non-zero index, first row index, first sub-block index), reduction information (row index of intermediate result, etc.), and so on. In an Operator Graph, operators convert a matrix by modifying the Matrix Metadata Set in order. After an Operator Graph has been iterated, Matrix Metadata Set will include all effects of operators to the original matrix cumulatively. A simple example of matrix metadata is shown on the lower part of \Cref{ExampleofFormatGeneration}. The red text represents where the metadata is added or modified. Take row\_indices and col\_indices as examples. They are added by the input matrix, recording the row and column indices of non-zeros, and operator BMT\_PAD further modifies them by adding an index of a zero element in a specific position (1, 1).


\subsection{Format Construction}

All arrays of a format are extracted from Matrix Metadata Set by choosing the metadata needed by the kernel (determined by kernel fragment detained in \Cref{KernelBuilder}). In the final format shown in \Cref{ExampleofFormatGeneration}, bmtb\_nz\_offsets, bmt\_row\_offsets, and bmtb\_bmt\_offsets record the indices of the first non-zero, sub-block, row in each BMT or BMTB. They are generated by operator BMT\_ROW\_BLOCK and BMTB\_ROW\_BLOCK, defining how the matrix is distributed to each thread and thread block. bmt\_sizes\_of\_bmtb is non-zero numbers of BMT in each BMTB, generated by BMT\_PAD. origin\_rows (generated by SORT) and bmt\_row\_offsets record the original row indices of intermediate results from non-zero multiplications, which are needed by reduction of GMEM\_ATOM\_RED (in vector \emph{y}).


\begin{figure}[htbp] 
\vspace{-0.5em}
\centering
\includegraphics[width=0.42\textwidth]{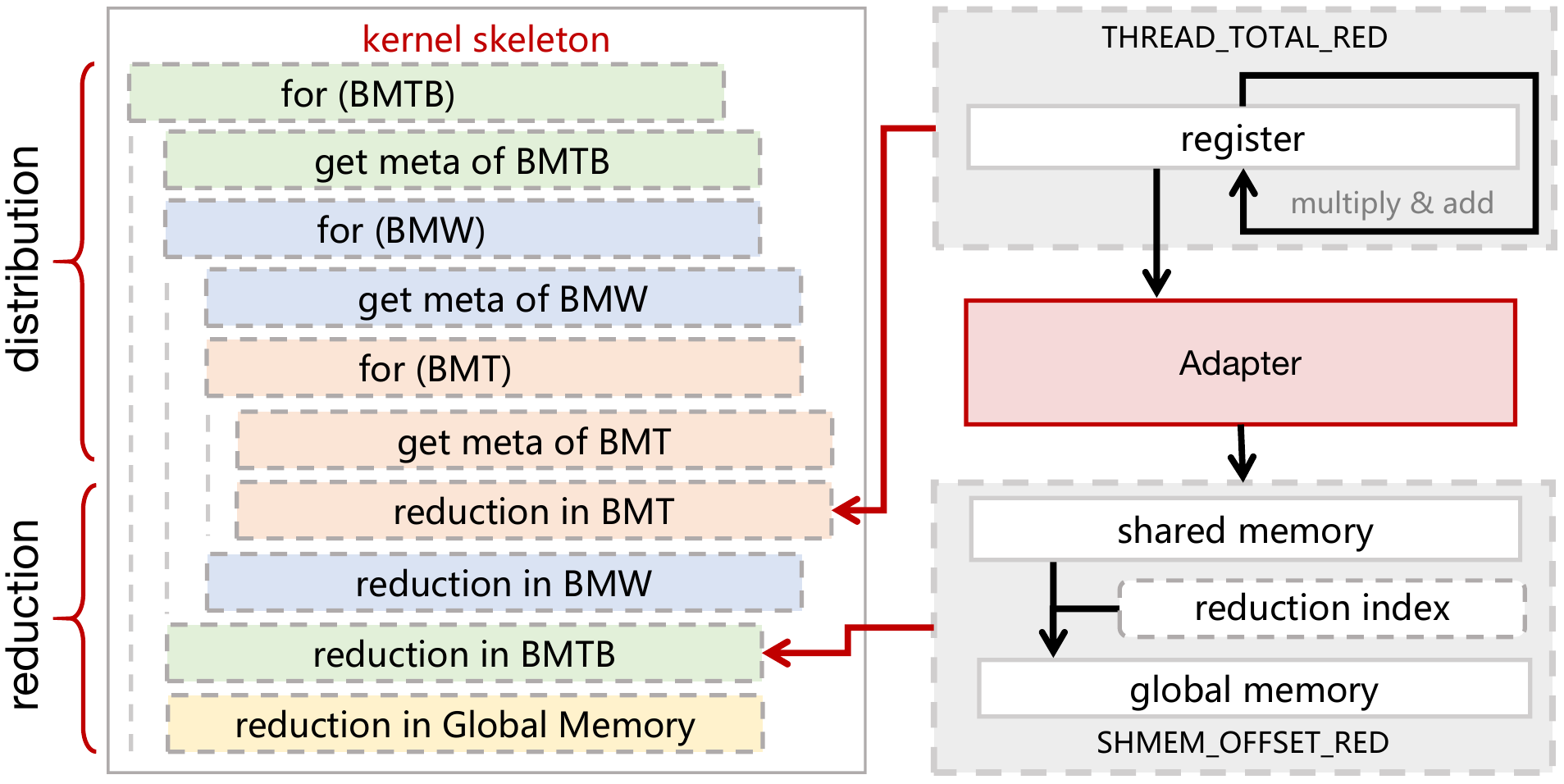} 
\vspace{-1em}
\caption{Example of kernel generation by splicing kernel fragment. The reduction strategies of this case are THREAD\_TOTAL\_RED and SHMEM\_OFFSET\_RED.}
\label{TemplateGenerationExample} 
\vspace{-0.5em}
\end{figure}
    
\subsection{Kernel Builder} \label{KernelBuilder}

\changed{The construction process of the SpMV kernel includes two parts}: 1) Distribution, mainly determined by the mapping stage. It gets metadata for each block in different levels of parallelism, which mainly includes information for task distribution and reduction strategy. 2) Reduction, mainly determined by the implementing stage. It multiplies the non-zeros of the matrix with the vector elements and reduces their results by row.

According to the commonality of SpMV programs, the template of Kernel Builder includes two key components: \emph{kernel skeleton} and \emph{kernel fragment}. The left of \Cref{TemplateGenerationExample} shows the kernel skeleton, which is the root symbol containing multiple nested loops. Each loop traverses blocks distributed to different levels of parallelism(thread block, warp, thread), including a series of slots for kernel fragments. Kernel fragments marked as ``get meta of BMX" read metadata arrays needed by other kernel fragments of the same loop, which constitute the format. It can be easily and automatically generated by analyzing data dependency. For the strategy to reduce current intermediate results, kernel fragments prefixed by ``reduction in" are determined by operators in the implementing stage. Non-orthogonal factors could appear in combinations of different reduction strategies. To solve this issue, special kernel fragments called \emph{Adapter} need to be pre-defined, \changed{which only includes several assignment expressions}. Shown on the right of \Cref{TemplateGenerationExample}, intermediate results from thread-level reduction (THREAD\_TOTAL\_RED) are further reduced in thread-block-level reduction (SHMEM\_OFFSET\_RED). The former reduction puts its output in the register group. The latter accepts input only in shared memory, which makes these two reduction strategies cannot be connected directly. An Adapter is needed to copy results from registers to shared memory in an accepted layout.

\begin{figure}[htbp]
\vspace{-0.5em}
\centering
\includegraphics[width=0.47\textwidth]{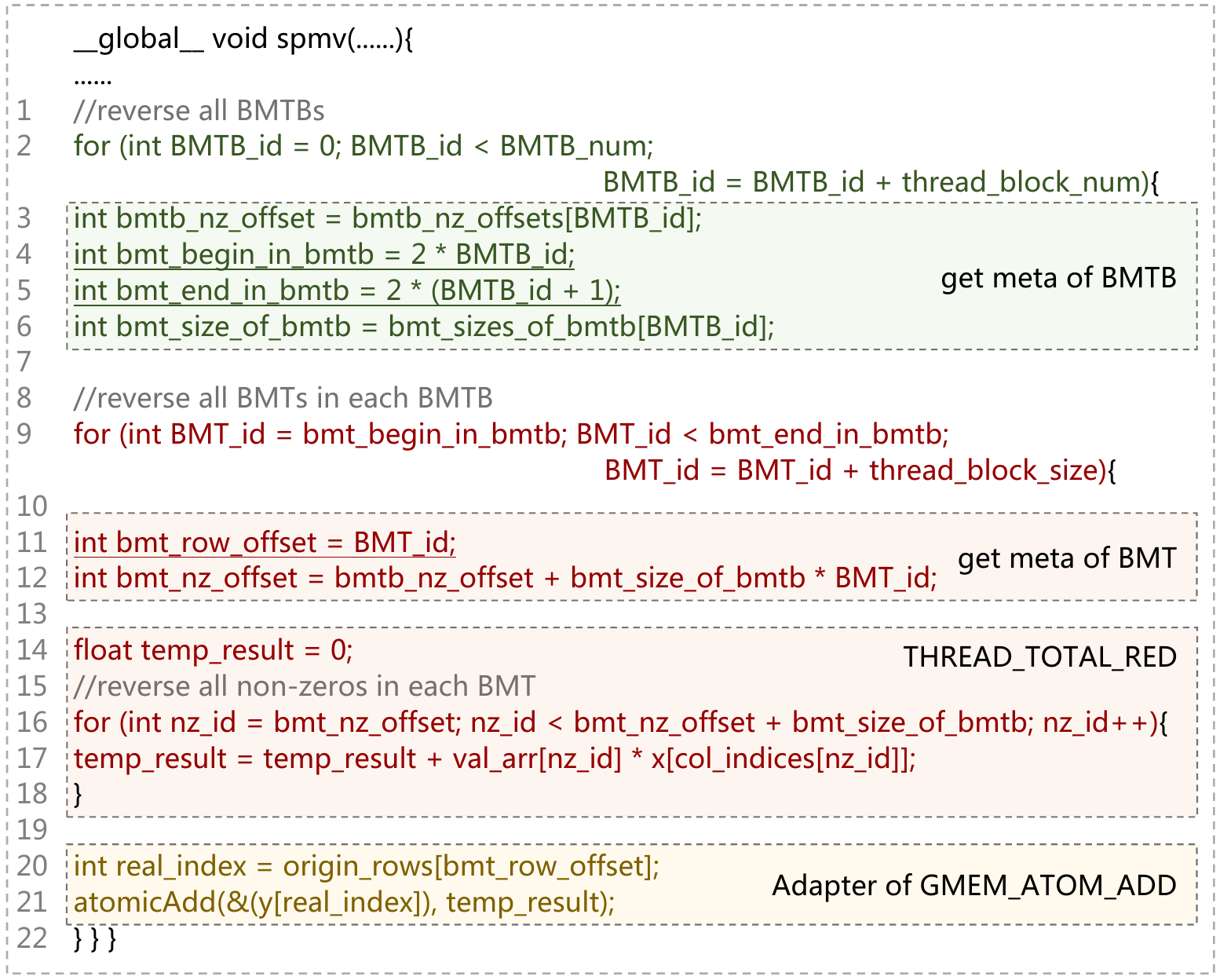} 
\vspace{-1em}
\caption{\changed{Example of generated kernel of Figure \ref{ExampleofFormatGeneration} after optimizing. Underlined codes are optimized by Model-Driven Format Compression.}}
\label{CodeGenerationExample} 
\vspace{-0.5em}
\end{figure}

\Cref{CodeGenerationExample} shows an example kernel of Operator Graph shown in \Cref{ExampleofFormatGeneration}. In this case, the whole matrix is divided into BMTBs and BMTs in the row direction. Each thread reduces its contents into one result. These results from threads are further reduced in the global memory. Lines 3-6 and 11-12 get format(metadata) arrays of BMTB and BMT. Lines 14-18 multiply non-zeros of each BMT by elements of vector \emph{x} and reduce them in one register represented by temp\_result. Lines 20-21 further reduce the intermediate result of each thread by atomic addition in the global memory. 

\subsection{Optimizer}

\begin{figure}[htbp] 
\vspace{-0.5em}
\centering
\includegraphics[width=0.46\textwidth]{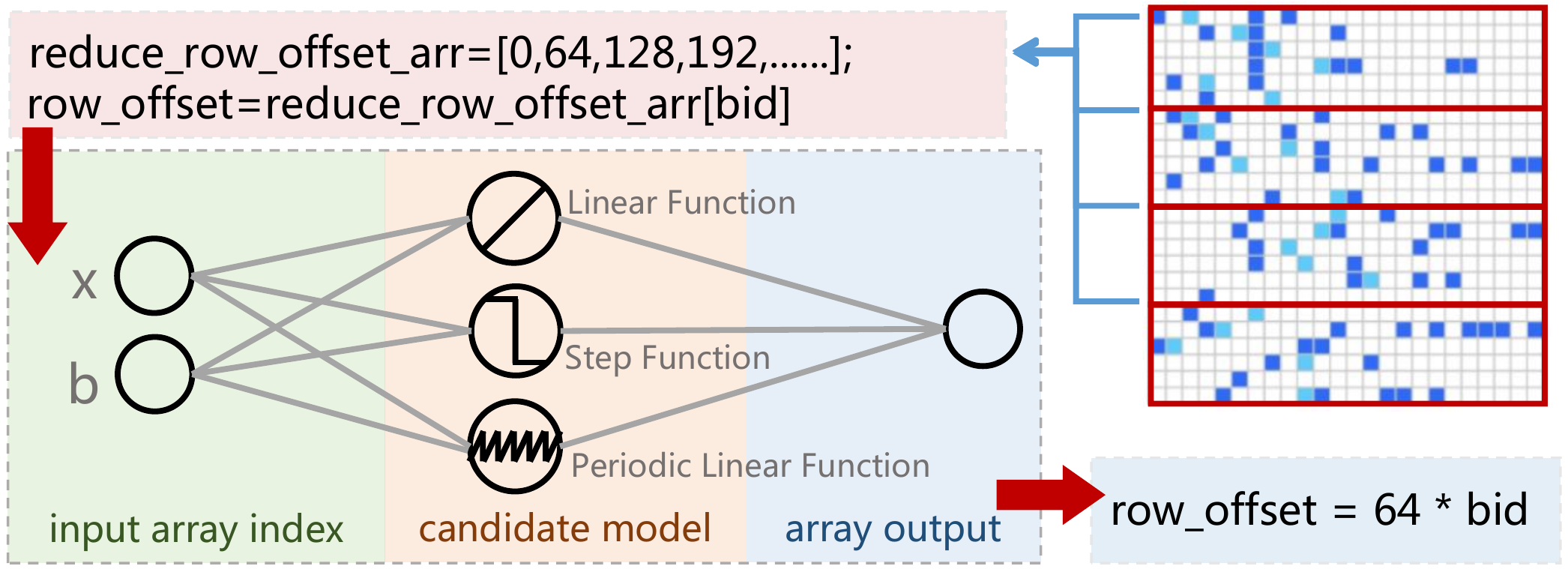} 
\vspace{-0.5em}
\caption{\changed{Example of memory access optimization for a format array.}}
\label{ExampleofMemoryAccessOptimization} 
\vspace{-0.5em}
\end{figure}

To improve kernel performance, Kernel Builder supports a series of optimizing strategies, such as removing unnecessary codes, combining multiple short-data-type arrays, etc. \changed{Since kernel optimizations have been well studied before, we leverage several state-of-the-art techniques in our tuning system. One prominent optimization is \emph{Model-Driven Format Compression} (derived from ~\cite{augustine2019generating}), which is especially efficient to memory-access optimization. It reduces the number of memory accesses by transforming array type data (in memory) to models and replacing memory access with calculation. 

As shown in \Cref{ExampleofMemoryAccessOptimization}, an operator named BMTB\_ROW\_DIV divides the matrix into row bands every 64 lines, and each row band is mapped to a thread block. It adds an array, named reduce\_row\_offsets, to the format, which includes the first-row index of BMTBs for thread-block-level reduction.} row\_offset can be calculated directly from the index of BMTB (row\_offset=64*bid), by fitting array index and value to a linear model, which makes global memory access (row\_offset=reduce\_row\_offsets[bid]) unnecessary. In manually written codes, programmers can naturally discover the regularity of data structures and directly write the optimized implementation. Because AlphaSparse is entirely automatic, we need to perform this optimization explicitly to achieve competitive performance with human-written codes. In addition to linear functions, other functions, such as step function and periodic linear function, are also supported. Users can also extend the hypothesis function. Unlike normal regression problems of data analysis, any errors in the model would cause incorrect SpMV implementation. To improve the success rate of this optimization, a small number of errors can be tolerated by adding \emph{if} statements to separately assign values for the specific array index that the model cannot fit. 

\Cref{CodeGenerationExample} includes example optimizations: Accesses of bmtb\_bmt\_offsets, bmt\_row\_offsets are eliminated by Model-Driven Compression; and the optimizer also eliminates the warp-level loop for a cleaner code structure.




\section{Search Engine} \label{DesignSpaceExploration}

\textit{Search Engine} drives AlphaSparse by enumerating Operator Graphs and choosing the best one of them. To deal with a huge search space (as a challenge detailed in \Cref{introduction}) consisting of the parameters and structure of the Operator graph, Search Engine provides multi-level search from coarse to fine. It exploits the experience of coarse-grained search to accelerate the fine-grained search by an ML model.


\subsection{Operator Graph Search}



The search strategy of the Search Engine includes three steps (levels). In the first step, graph structures are enumerated by randomly choosing empty operators and connecting them at the end of the existing Operator Graph. The second step searches node (operator) parameters in a coarse-grained grid and gets the performance of Operator Graphs by directly running corresponding SpMV programs. In the third step, the test results from step two are further interpolated to a fine-grained parameter grid by an ML model. We do not directly do the fine-grained search because the overhead of running SpMV programs is extremely high, even occupying almost all the searching overhead. In comparison, the overhead of the ML model is negligible. To further and reasonably reduce the executions of SpMV programs, the first two steps could be terminated early by simulated annealing. Moreover, we also limit the search time to no more than 8 hours, as a mandatory termination condition, according to our experience.

According to our practice, XGBoost\cite{chen2016xgboost} performs very well in interpolation, which is also confirmed to be practical by TVM~\cite{chen2018tvm}. It achieves a mean absolute deviation of 5\%, which is even less than the performance volatility of GPU. Because of the memory hierarchy of the architectures, we speculate that the cost model of memory-bound programs includes linear decision boundaries, which fits a tree-type model. The third step significantly reduces the overhead of the parameter search. Assume there is an Operator Graph with $q$ parameters. Reducing the search step size by half would increase search space by $2^q$ times, finally increasing the search time from several hours to several weeks. XGBoost can achieve the same effect by incurring relatively negligible overhead.

\subsection{Pruning} \label{Pruning}



Although AlphaSparse provides a three-level search to accelerate the searching process, the overhead from the first two steps is still expensive because of the huge search space of AlphaSparse. So, in addition to coarse-grained parameter search and simulated annealing, more pruning strategies are needed. 

\noindent\textbf{Pruning the search of the parameter}. Parameters indicate quantifiable details of an operator. The biggest challenge is the array type parameter. For example, ROW\_DIV includes an array type parameter containing the positions where the matrix is divided in the row direction. Assuming the input matrix has $10^5$ rows, the search space size of just this single parameter is $10^5!$, which is impossible to grasp. One or more parameter discretization strategies are included in each operator to handle array type parameters. \emph{Parameter discretization strategies} can reduce the parameter space, especially spaces of array-type parameters. In this case, a parameter discretization strategy named DIV\_IN\_ROW\_LEN\_MUTATION can be used to divide the matrix where row length mutates. It converts the array type parameter to just several integer parameters describing the degree of such mutation, which can be easily enumerated. 

\noindent\textbf{Pruning the search of the graph structure}. Pruning strategies for graph structure are added when we find operators are unnecessary for specific matrix sparsity patterns. For example, matrices with short rows do not need to try operators for long row reduction. Users can add their pruning strategies. AlphaSparse provides a ban list for pruned operators, according to already existing operators of graph and sparsity patterns of input matrices.


\section{Evaluation}\label{evaluation}

Our evaluation shows that AlphaSparse provides the highest overall performance among the most advanced artificial formats and the up-to-date implementation of traditional auto-tuning. 




\begin{figure*}[htbp] 
\vspace{-0.5em}
\centering
\includegraphics[width=0.85\textwidth]{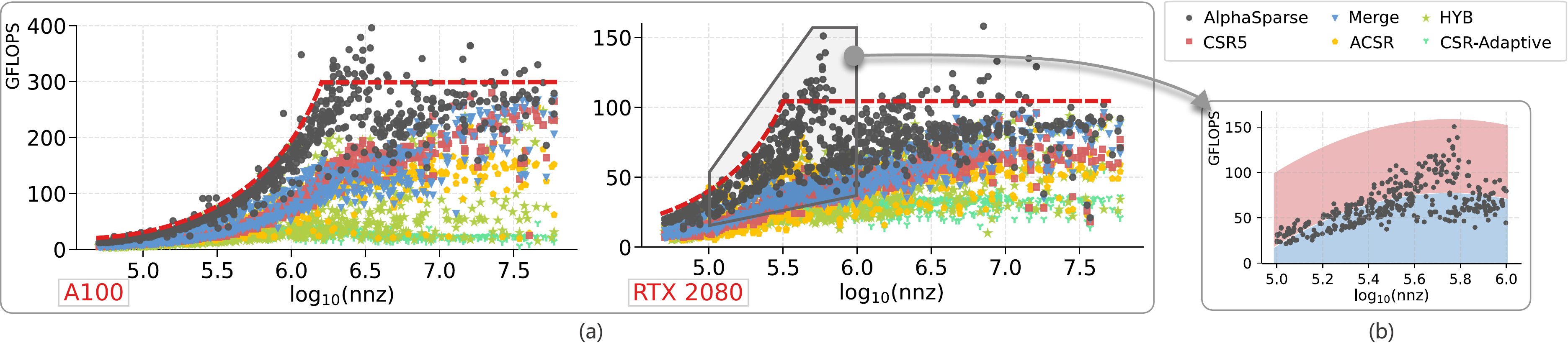} 
\vspace{-1em}
\caption{SpMV overall performance of matrices with different sizes. (a) All test results on RTX 2080 and A100, while the red dashed lines show a trend of \changed{achieved highest performances}. (b) Parts of the AlphaSparse results on RTX 2080. The region colored by red includes cases providing higher performance in specific matrix size, while the other is colored by blue.}
\label{TestOverview} 
\vspace{-0.5em}
\end{figure*}


\subsection{Experimental Setup}


\noindent\textbf{Platform}. The experiments are conducted on NVIDIA A100 and RTX 2080. The former is based on Ampere architecture, with 6912 CUDA cores, 40GB HBM2 memory \changed{(1.5TB/s)}, and 19.49 TFLOPS peak performance. The latter is based on Turing architecture, with 2944 CUDA cores, 8GB GDDR6 memory \changed{(448GB/s)}, 10.07 TFLOPS peak performance. We use single-precision for floating-point values in experiments.

\noindent\textbf{Testset}. The experiment includes 843 matrices (most of them are irregular) from SuiteSparse Matrix Collection~\cite{10.1145/2049662.2049663} whose features satisfy the three conditions: 1) row number is larger than 9K, 2) number of non-zeros is between 50K and 60M, 3) no empty rows \footnote{Our prototype has not included operators to handle empty rows.}. We ignore matrices with extremely large sizes because they are difficult to grasp. Small matrices are also ignored because they are not suitable for GPU.

\subsection{Baselines}

\changed{The baselines are classified into three kinds according to the degree of coupling with SpMV. {\bf Artificial format} represents the special library achieved by hand. {\bf Format selector} represents the traditional auto-tuning framework. {\bf Tensor algebra compiler} represents the more general compiler that considers SpMV as one of many objects.}

\noindent\textbf{Artificial format}. To compare with artificial formats, we choose several popular state-of-the-art formats with high performance and irregularity-specific design as follows: 1) \texttt{ACSR}~\cite{ashari2014fast}, implemented by us because so far we have not found its high-quality implementation. 2) \texttt{CSR-Adaptive}~\cite{daga2015structural}, from ViennaCL 1.7.1~\cite{rupp2010viennacl, viennacl.org}. 3) \texttt{CSR5}~\cite{liu2015csr5}\footnote{\url{https://github.com/weifengliu-ssslab/Benchmark_SpMV_using_CSR5}}, 4) \texttt{Merge-based CSR(Merge)}~\cite{merrill2016merge}\footnote{\url{https://github.com/dumerrill/merge-spmv}}. 5) \texttt{HYB}, from cuSPARSE 9.2.

\noindent\textbf{Format selector}. It is unrealistic to fairly compare AlphaSparse with the traditional auto-tuning philosophy based on format selection. The most state-of-the-art auto-tuners, Zhao et al.~\cite{zhao2018bridging}, SMAT(ER)~\cite{10.1145/3218823} and clSpMV, have historical limitations: 1) They contain only out-of-date formats, which sometimes cannot handle irregularity and cannot take advantage of new GPU features. 2) They have not been actively maintained for a long time. \footnote{clSpMV has not been maintained for seven years, and we could not deploy it on our platform due to the error appearing. We believe AlphaSparse can outperform clSpMV because AlphaSparse gains better performance than \texttt{ACSR} which outperforms clSpMV.}  For a reasonable comparison, we implement a Perfect Format Selector (PFS) as a representative to the up-to-date auto-tuner as the baseline. 

As a performance-first auto-tuner, PFS does not rely on probabilistic models for format selection. To achieve the highest accuracy(100\%), PFS can certainly select the best formats by directly running SpMV of all candidate formats. For an up-to-date implementation, PFS consists of five aforementioned state-of-the-art formats: \texttt{ACSR}, \texttt{CSR-Adaptive}, \texttt{CSR5}, \texttt{Merge}, \texttt{HYB}~\cite{bell2009implementing}; three root formats from the widely-used cuSPARSE library: \texttt{ELL} (from v9.2), \texttt{COO}, \texttt{CSR} (from v11.6); and two derived formats: \texttt{SELL}, \texttt{row-grouped CSR}, for a comprehensive comparison.

\changed{\noindent\textbf{Tensor algebra compiler}. Compiler focuses more on code-level optimizations instead of algorithm-level designs. For a more sufficient comparison, we add TACO~\cite{kjolstad2017tensor}\footnote{\url{https://github.com/tensor-compiler/taco }we use CUDA code fully automatically generated by it.} as a baseline of the tensor algebra compiler.}

\subsection{Performance Comparison with Artificial Formats}



\Cref{TestOverview}a shows the overall performance of AlphaSparse and state-of-the-art formats in 843 matrices. The x-axes are matrix size and we use floating point operations per second(GFLOPS) to represent performance as the y-axes. AlphaSparse achieves the highest performance among all artificial formats. On A100, AlphaSparse achieves an average $3.2\times$ speedup and the maximum $22.2\times$ speedup (in matrix \inds{TSOPF\_RS\_b300\_c2}) over all artificial formats. In particular, it achieves average $2.3\times$, $5.7\times$, $2.0\times$, $2.0\times$ and $3.9\times$ speedup over \texttt{ACSR}, \texttt{CSR-Adaptive}, \texttt{CSR5}, \texttt{Merge} and \texttt{HYB}, respectively. AlphaSparse outperforms \texttt{Merge}, \texttt{ACSR}, \texttt{CSR-Adaptive} and \texttt{CSR5} in all 843 matrices, while it outperforms \texttt{HYB} in 841 matrices (because AlphaSparse has not included the matrix decomposition strategy of \texttt{HYB}). On RTX 2080, AlphaSparse achieves an average $2.0\times$ speedup and the maximum $8.3\times$ speedup (in matrix \inds{TSOPF\_RS\_b2052\_c1}). In particular, it achieves average $2.0\times$, $2.3\times$, $2.0\times$, $1.7\times$ and $2.4\times$ speedup over \texttt{ACSR}, \texttt{CSR-Adaptive}, \texttt{CSR5}, \texttt{Merge} and \texttt{HYB}, respectively.

\texttt{Merge} and \texttt{CSR5} provide the highest overall performance among all artificial formats, because they benefit from thread-level load balance by allocating a balanced number of non-zeros or rows to each thread. The overall performance of \texttt{CSR-Adaptive} is the lowest. It performs well in relatively small matrices by achieving higher parallelism. However, it suffers from giving up the reduction in registers, making it perform the worst on remaining matrices. \texttt{ACSR} and \texttt{HYB} are based on matrix decomposition, providing mild performance. 

In \Cref{TestOverview}a, maximum performances of AlphaSparse in each matrix size make up \changed{a trend of} flat-tail shape, represented by red dashed lines. As a memory-bound program, the SpMV performance can be raised by the increasing occupy of memory bandwidth when the matrix size is not too large. When the memory bandwidth is sufficiently used, the performance will not further increase~\cite{williams2009roofline}. In our evaluation, only AlphaSparse approaches this trend. 



To show how input matrices affect performances, we take samples of RTX 2080 test results and divide them into two parts in \Cref{TestOverview}b. We choose this range of results because it shows \changed{clear upper and lower borders and makes us easy to split in the middle of them.} Although these two parts of cases correspond to the same matrix sizes, the performance of the upper part (red) is up to $5.0\times$ (average $1.4\times$) higher than the lower. According to our further observation, we suspect that the two matrix features cause this performance gap. One is average row length($\frac{nnz}{n}$), which in the upper part is $1.9\times$ higher than the lower. We speculate that a higher average row length \changed{improves performance by increasing the ratio of calculation to memory access and decreasing the proportion of reduction operations (which require synchronization) in the SpMV program.} The other is row variance(degree of regularity, $\frac{\sum{(row\_len-\frac{nnz}{n})^{2}}}{n}$), which in the upper part is $20\times$ lower than the lower. Lower regularity can usually achieve higher reduction performance, better load balance, and less computation waste.

\begin{figure}[htbp] 
\vspace{-0.5em}

\centering
\includegraphics[width=0.45\textwidth]{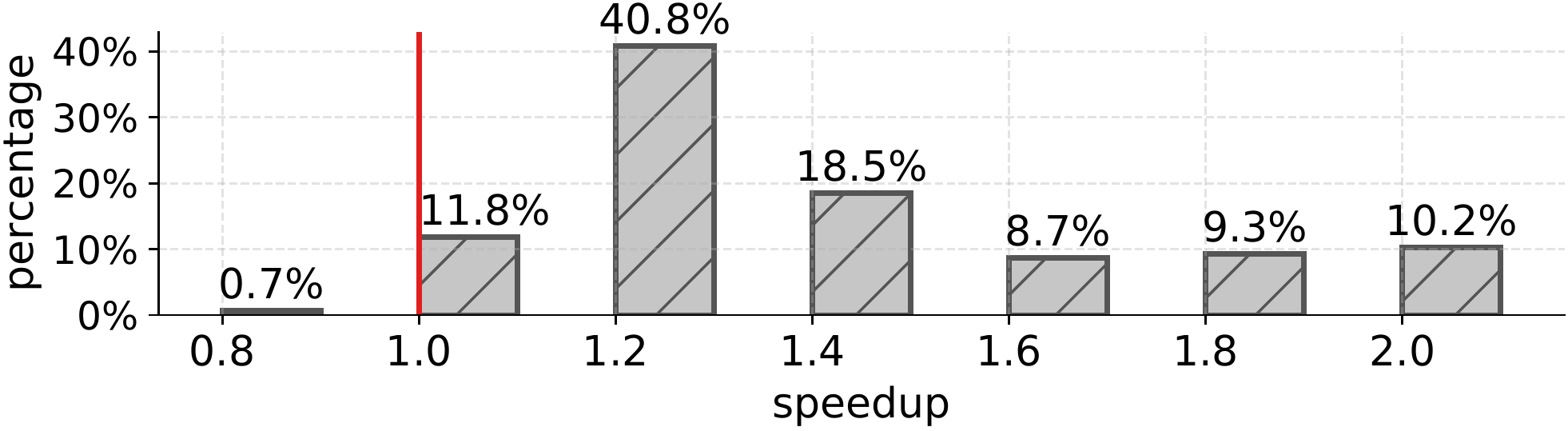} 
\vspace{-1em}
\caption{The frequency distribution of AlphaSparse's speedup over PFS on A100.}
\label{TheFrequencyDistributionofMax} 
\vspace{-0.5em}
\end{figure}

\begin{figure}[htbp] 
\vspace{-0.5em}
\centering
\scriptsize
\begin{tabular}{cc}
\includegraphics[width=0.23\textwidth]{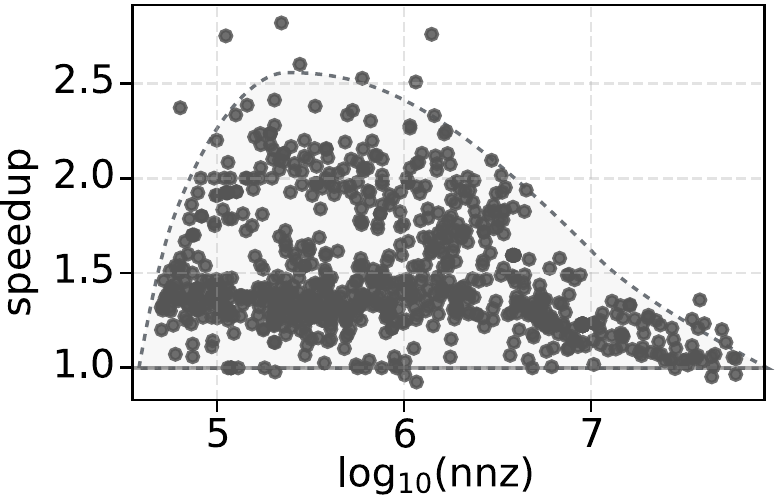} & \includegraphics[width=0.212\textwidth]{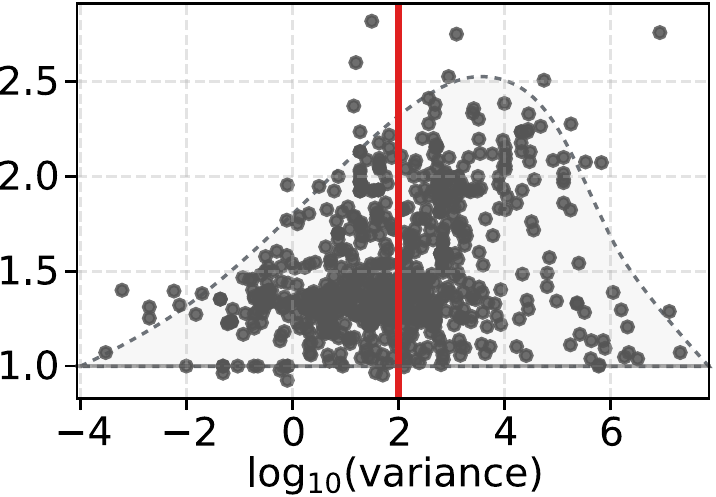}\\
(a) & (b)
\end{tabular}
\vspace{-1em}
\caption{\changed{Speedups of AlphaSparse over PFS corresponding to (a) matrix sizes and (b) variances of row lengths on A100.}}
\label{SpeedUptoBest} 
\vspace{-0.5em}
\end{figure}

\subsection{Performance Comparison with Format Selector}


\Cref{TheFrequencyDistributionofMax} illustrates the frequency distribution of AlphaSparse's speedup over PFS on A100. In 99.3\% cases, the performance of AlphaSparse is higher. In the remaining 0.7\% matrices, AlphaSparse performs worse because some design strategies of formats in PFS are not included in AlphaSparse (detailed in \Cref{Limitation}). Most (40.8\%) cases achieve the speedup between $1.2\times$ and $1.4\times$. 

\Cref{SpeedUptoBest} further demonstrates speedups of AlphaSparse over PFS along with matrix sizes and row variances(to show the influences of the irregular sparsity). \changed{\Cref{SpeedUptoBest}a shows impressive speedups can be achieved in cases where the matrix fits into the 40 MB L2 cache of the A100, and large matrices ($\ge10^7$ non-zeros) provide lower speedups.} In \Cref{SpeedUptoBest}b, the red line shows the boundary of the regularity and irregularity ($10^2$ as mentioned). The peak of speedup is $2.7\times$, appearing in the middle degree of matrix size and irregularity, which shows the fine-grained trade-off provided by Operator Graph is suitable for moderate sparsity patterns. On the contrary, designs of most artificial formats are based on human observations of specific extreme sparsity patterns from matrices such as \inds{Webbase}, \inds{mip1}, \inds{FullChip}, as shown in their papers. They ignore matrices with moderate sparsity patterns. Moreover, we find irregular matrices benefit more from AlphaSparse: the average speedup is $1.4\times$ for regular sparsity, while for irregular sparsity, the average speedup goes up to $1.6\times$.


\begin{figure}[htbp] 
\vspace{-0.5em}
\centering
\scriptsize
\begin{tabular}{cc}
\includegraphics[width=0.234\textwidth]{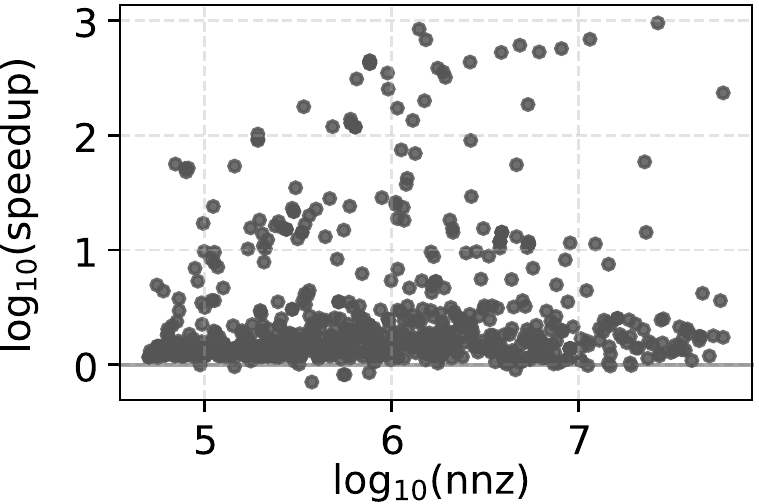} & \includegraphics[width=0.221\textwidth]{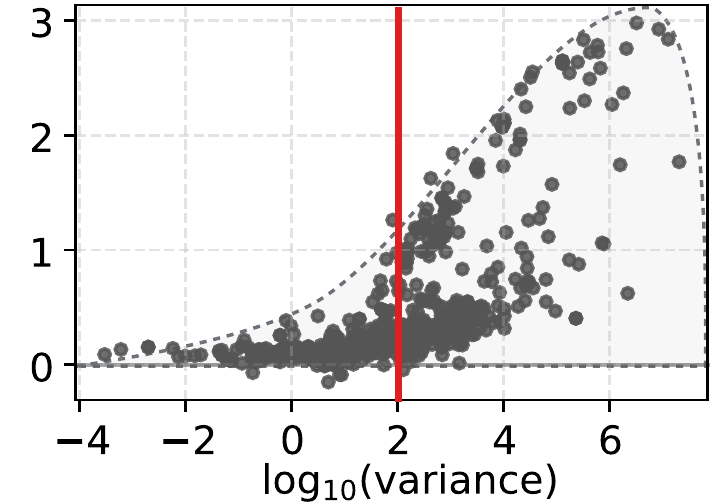}\\
(a) & (b)
\end{tabular}
\vspace{-1em}
\caption{\changed{Speedups of AlphaSparse over TACO corresponding to (a) matrix sizes and (b) variances of row lengths on A100.}}
\label{SpeedUptoTACO} 
\vspace{-0.5em}
\end{figure}

\begin{figure}[htbp] 
\vspace{-1em}
\centering
\includegraphics[width=0.43\textwidth]{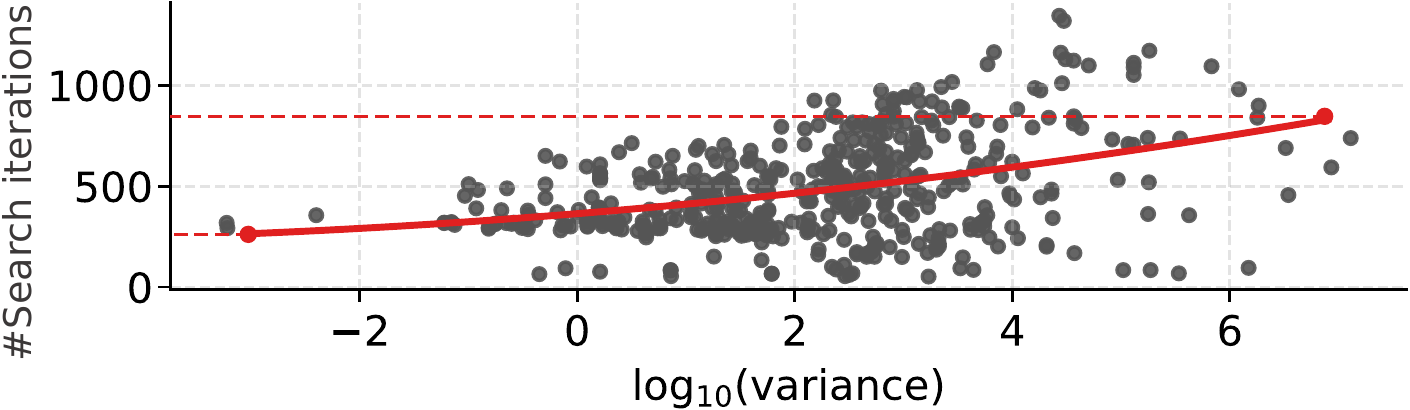}
\vspace{-1em}
\caption{Numbers of searching iterations along with row variances on A100.}
\label{A100SearchIterNum} 
\vspace{-0.5em}
\end{figure}

\begin{table}[htbp]
\vspace{-0.5em}
\caption{\changed{Search time and performances with and without pruning on A100}}
\vspace{-0.5em}
\scriptsize
\centering
\begin{tabular}{c|c|c|c|c}
\toprule
\multirow{2}{*}{Matrix} & \multicolumn{2}{c|}{\textbf{Search Time (hour)}} & \multicolumn{2}{c}{\textbf{Performance (GFLOPS)}} \\
\cline{2-5}
& no pruning & pruning & no pruning & pruning \\
\midrule
\inds{pdb1HYS} &  \multirow{13}{*}{8.0} & 3.1 & 273.4 & 303.2 \\
\inds{windtunnel\_evap3d} &  & 2.3 & 286.1 & 343.3 \\
\inds{consph} &  & 1.9 & 339.8 & 356.0 \\
\inds{Ga41As41H72} &  & 3.4 & 193.6 & 242.1 \\
\inds{Si41Ge41H72} &  & 4.8 & 175.1 & 236.9 \\
\inds{ASIC\_680k} &  & 0.9 & 121.8 & 169.6 \\
\inds{mip1} &  & 4.9 & 227.0 & 226.0 \\
\inds{Rucci1} &  & 1.9 & 218.9 & 223.7 \\
\inds{boyd2} &  & 2.4 & 61.3 & 80.2 \\
\inds{rajat31} &  & 5.1 & 189.2 & 226.0 \\
\inds{transient} &  & 3.0 & 127.7 & 153.0 \\
\inds{ins2} &  & 3.4 & 101.9 & 152.8 \\
\inds{bone010} &  & 3.3 & 193.2 & 235.4 \\
\hline
\textbf{Average}& 8.0 & 3.2 & 198.6 & 231.0 \\
\bottomrule
\end{tabular}
\label{PruningTest}
\vspace{-1.5em}
\end{table}

\subsection{Performance Comparison with TACO}

\changed{AlphaSparse greatly outperforms TACO. On A100, AlphaSparse achieves an 18.1$\times$ average speedup and the maximum 950.8$\times$ speedup over TACO. As shown in \Cref{SpeedUptoTACO}a, speedups are insensitive to matrix sizes, unlike PFS. \Cref{SpeedUptoTACO}b shows the peak of speedup appearing in highly irregular matrices. Two reasons cause its relatively lower performance. The first reason is that TACO is not tailored for SpMV. Its three key features, index compression, loop optimization, and automatic parallelism, only target general sparse problems. None of them can handle problems brought by SpMV, especially the irregularity. The second reason is that TACO lacks the utilization of GPU features, which even lacks competitiveness with human-designed programs.}

\subsection{Searching Overhead}

Since the first two search steps occupy almost all the searching overhead (as mentioned in \Cref{DesignSpaceExploration}), we use the number of iterations in the first two steps to represent the performance of search strategies. \Cref{A100SearchIterNum} shows search iterations along with degrees of matrix irregularity (so-called row variances). The regression line of test results illustrates a positive correlation between the search overhead and matrix irregularity: regular matrices need $3.5\times$ fewer iterations than highly irregular matrices. These prove that our pruning rules significantly reduce search overhead by ignoring operators for the irregularity when the input matrix is regular.


\changed{\Cref{PruningTest} shows how pruning strategies affect search time and performances of AlphaSparse. We record test results before\footnote{\changed{We remove the simulated annealing and other pre-defined pruning strategies (shown in \Cref{Pruning}) in the baseline of no pruning.}} and after pruning in 13 popular matrices evaluated from published researches. Pruning strategies reduce search time by 2.5$\times$ on average. Because pruning strategies include high-quality human experience, they eliminate unnecessary enumerations and make the Search Engine focus on areas of the design space that are highly likely to find high-performance formats in limited search time. Pruning strategies also improve performance by 1.2$\times$ on average. Compared with existing offline auto-tuners, such as PATUS (8 hours)~\cite{christen2011patus}, SDSL ($\ge$33 hours)~\cite{henretty2013stencil}, Halide (2 hours-2 days)~\cite{ragan2013halide}, PARTANS (2.5 hours-32 days)~\cite{lutz2013partans}, the search time of AlphaSparse is competitive.}

\begin{figure}[htbp] 
\vspace{-0.5em}
\centering
\scriptsize
\begin{tabular}{c}
\includegraphics[width=0.47\textwidth]{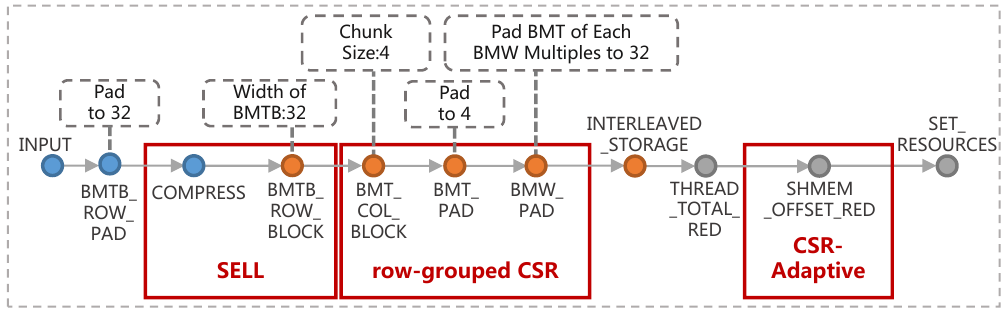}\\
(a)\\
\includegraphics[width=0.45\textwidth]{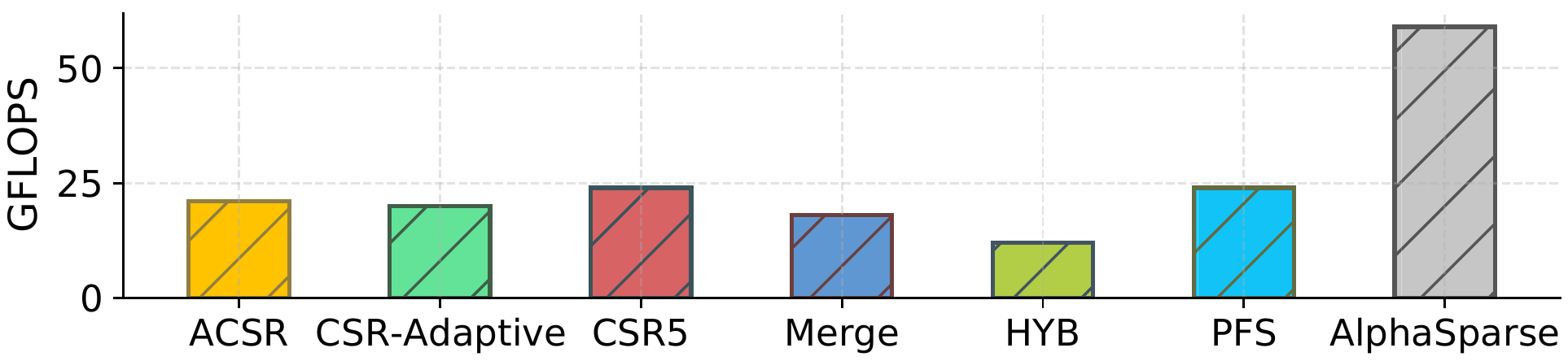}\vspace{-0.5em}\\
(b)\\
\includegraphics[width=0.45\textwidth]{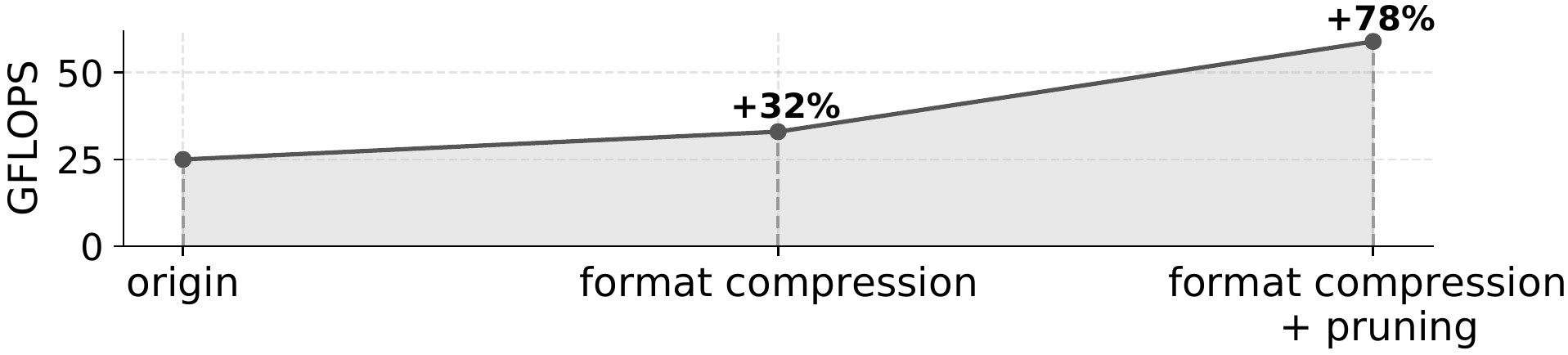}\vspace{-1em}\\
(c)\\
\end{tabular}
\vspace{-1em}
\caption{An example of matrix \inds{scfxm1-2r} on A100. (a) a snapshot of its Operator Graph, (b) the performance comparison with artificial formats and \changed{(c) performance improvements achieved by two key optimizations of AlphaSparse}.}
\label{AnOperatorGraph} 
\vspace{-0.5em}
\end{figure}

\subsection{Creative Capability of AlphaSparse}

Creating new machine-designed formats is the main driver of performance improvement. From our statistics, in 73.1\% of test cases, AlphaSparse outperforms all counterparts by creating machine-designed formats which are not covered by its source formats (as referenced in  \Cref{TheExampleofFormatDesignNode}). In 16.5\% of cases providing new formats, the branches appear in Operator Graphs, which means AlphaSparse designs different formats and corresponding kernel implementations for different parts of original matrices.

\Cref{AnOperatorGraph}a shows an example Operator Graph of an new format generated by AlphaSparse for matrix \inds{scfxm1-2r}. It mainly includes the thread-block-level blocking strategy from \texttt{SELL}, the thread-level blocking strategy from \texttt{row-grouped CSR}, and the reduction strategy from \texttt{CSR-Adaptive}. \changed{Finally, as shown in \Cref{AnOperatorGraph}b, it achieves $2.7\times$ speedup (which is the highest) over PFS and state-of-the-art artificial formats. Appropriate trade-offs between different design strategies achieve high performance. Compared with the source formats, for this matrix, the machine-designed format avoids high padding rate of \texttt{SELL}, inefficient global memory reduction of \texttt{row-grouped CSR}, ignorance of thread-level reduction in \texttt{CSR-Adaptive}, and benefits from regular row block indices of \texttt{SELL}, low padding rate of \texttt{row-grouped CSR}, efficient shared memory reduction of \texttt{CSR-Adaptive}. In terms of its state-of-the-art counterparts, expensive strategies, such as binning of \texttt{ACSR} and blocking for load balancing of \texttt{CSR5} and \texttt{Merge-based CSR}, are unnecessary because the matrix is not too irregular. Moreover, \texttt{HYB} includes a large, inefficient \texttt{COO} component in this matrix, which makes it also worse than the machine-designed format. \Cref{AnOperatorGraph}c shows a 32\% performance improvement is brought by Model-Driven Format Compression, and pruning strategies bring a further 78\% performance improvement.}



\subsection{Limitation}\label{Limitation}

In AlphaSparse, the lack of operators is the main reason causing slightly lower performance on specific matrices. \changed{A representative case is matrix \inds{GL7d19}. Its best artificial format is HYB, which performs even better than machine-designed formats from AlphaSparse. In this matrix, the lengths of almost all rows are relatively balanced, except for a few rows that are several times longer. The matrix decomposition strategy of HYB is very suitable for this sparsity pattern, but the current AlphaSparse has not included this strategy.}

\changed{In addition to HYB-like decomposition, two} popular categories of operator have not been included: operators for local densities~\cite{langr2012adaptive, choi2010model, karakasis2009comparative}, diagonal patterns~\cite{weber2013efficient, kourtis2011csx}. They separate the regular parts of the matrix and handle them exclusively to achieve high performance. However, they only cover a small number of matrices. Our prototype implementation has not considered them, but they will be considered for more complete support in the future.

\changed{Currently, the proof-of-concept AlphaSparse only supports CUDA. However, it can be extended to other platforms by implementing new tailored operators. Users only need to define how the operator modifies metadata and occasionally need to define kernel fragments.}

\section{Related Works}\label{relatedworks}

\noindent \textbf{Auto-tuners}. Auto-tuners have proven to be a successful performance tuning approach, represented by ATLAS \cite{whaley1998automatically}, FFTW \cite{frigo2005design}, SPIRAL \cite{puschel2005spiral}, and OSKI \cite{vuduc2005oski}, for the increasingly diverse and complicated computer architecture designs.
For sparse linear algebra, SMAT~\cite{li2013smat}, clSpMV~\cite{su2012clspmv}, TileSpMV~\cite{niu2021tilespmv} Naser Sedaghati et al.~\cite{sedaghati2015automatic} and \texttt{CSX}~\cite{kourtis2011csx} select the best artificial format and SpMV implementation for the given matrix; while IA-SpGEMM \cite{xieia2019} selects the best formats for SpGEMM.
TVM \cite{chen2018tvm} and Ansor \cite{zhen2020ansor} are auto-tuner for dense tensor calculation by automatically generating code structure and selecting the best corresponding parameters. COGENT \cite{kim2019code} provides high performance tensor contractions on GPU. CASpMV~\cite{xiao2019caspmv} include auto-tuner for matrix partition on the Sunway. Some general auto-tuners, ATF \cite{rasch2019atf}, OpenTuner \cite{ansel2014opentuner}, CLTune \cite{nugteren2015cltune}, Optuna \cite{akiba2019optuna}, mNM~\cite{balaprakash2011can}, Muthu et al. \cite{baskaran2008compiler}, Tiwari et al.~\cite{tiwari2009scalable}, Rigel \cite{sreenivasan2019framework} and SMAC3 \cite{hutter2011sequential}, have been designed to ease the designing effort of an auto-tuner and target in a broader scope.  AlphaSparse is not limited by selecting among artificial formats, kernel implementations, parameters, and it is able to create SpMV code and break through the limits of human design.


\noindent \textbf{Artificial format and kernel design}. To improve the performance of SpMV, a dozens of formats \cite{langr2015evaluation, filippone2017sparse} have been proposed. State-of-the-art formats are derived from several base formats. \texttt{ALIGNED\_COO}~\cite{shah2012efficient}, \texttt{SCOO}~\cite{dang2013cuda}, \texttt{BRO-COO}~\cite{tang2014family}, \texttt{BCOO}~\cite{yan2014yaspmv} are derived from \texttt{COO}. \texttt{ICSR}~\cite{yang2012improved}, \texttt{CSR-Adaptive}~\cite{greathouse2014efficient}, \texttt{ACSR}~\cite{ashari2014fast}, \texttt{CSR5}~\cite{liu2015csr5}, \texttt{LightSpMV}~\cite{liu2015lightspmv} are derived from \texttt{CSR}. \texttt{ELL-R}~\cite{vazquez2011new}, \texttt{AdELL}~\cite{maggioni2013adell}, \texttt{JAD}~\cite{li2013gpu} are derived from \texttt{ELL}. \texttt{HYB}~\cite{bell2008efficient}, \texttt{HDC}~\cite{yang2014optimization} and \texttt{HEC}\cite{liu2012sparse} are hybrid formats. These artificial formats are designed by human according to their observations, AlphaSparse can automatically creates formats without human intervene.

\noindent \textbf{Code generation}. 
TVM \cite{chen2018tvm} is a template-based machine code generator for dense tensor calculation. TACO \cite{kjolstad2017tensor} can handle high-order sparse tensor calculation by compressing the index of each dimension. LL \cite{arnoldspecifying2010} is a DSL to define matrix format and its SpMV kernel. AlphaSparse provides a graph-based expression for generating of both format and kernel.

\section{Conclusion and Future Work}\label{conclusionandfuturework}

We present AlphaSparse, a fully automatic SpMV code designer, that generates outperforming format and kernel directly from an input sparse matrix. It unifies the modeling of format and kernel implementation and achieves an impressive speedup of up to 22.2$\times$ over state-of-the-art human-designed formats on the NVIDIA GPU.
We will examine advanced search strategies from existing research~\cite{chen2018tvm, vasilache2018tensor, chou2020automatic}, and add efficient format conversion routines in the future.

\section*{Acknowledgements}

This project is supported by National Natural Science Foundation of China under Grant No. T2125013, 62032023, 61972377, 61702483, and international partnership program of Chinese Academy of Sciences, Grant No. 171111KYSB20180011, and China National Postdoctoral Program for Innovative Talents BX2021320.


\bibliographystyle{IEEEtran}
\bibliography{IEEEabrv, bibfile}



\end{document}


\setcopyright{relax}
\settopmatter{printfolios=false, printccs=false, printacmref=false}
\copyrightyear{2022}
\acmYear{2022}
\acmConference[SC '22]{The International Conference for High Performance Computing, Networking, Storage, and Analysis}{November 13--18, 2022}{Dallas, TX, USA}
\acmBooktitle{The International Conference for High Performance Computing, Networking, Storage, and Analysis (SC '22), November 13--18, 2022, Dallas, TX, USA}
\acmPrice{}
\acmDOI{}
\acmISBN{}

\sloppy
\maketitle

\renewcommand{\shortauthors}{ Du, et al. }

\section*{Summary of the experiments reported}
AlphaSparse achieves up to 22.2 times (3.2 times on average) speedup over state-of-the-art formats and 2.8 times (1.5 times on average) over the up-to-date implementation of the traditional auto-tuning philosophy. Users need to input a Matrix Market file of a sparse matrix, and AlphaSparse will output the Operator Graph and kernel implementation. The experiments are conducted on NVIDIA A100 and RTX 2080. We use single-precision for floating-point values in experiments.

\section*{Download data of test matrices}

In our evaluation, all input matrices are from SuiteSparse Matrix Collection (https://sparse.tamu.edu). Two ways are provided to download data as follows. 

\noindent\textbf{Directly download from websites.} Go to the websites of SuiteSparse Matrix Collection, click the link of specific matrix name, and click the download link named ``Matrix Market''. The downloaded file is zipped. By extracting the file, a ``.mtx'' file can be gotten.

\noindent\textbf{Use the python interface.} Install the \textit{ssgetpy} Python module. Run ``import ssgetpy'' and type ``help(ssgetpy)'' to get a detailed help message on using ssgetpy to search and download sparse matrices. We have provided a python script named ``get\_data\_set\_from\_UF.py'' to download all the needed matrices. The variable ``UF\_DIR'' is the destination of downloaded data.

\section*{Reproduce AlphaSparse}

AlphaSparse is implemented by C++11 codes. The automatically output SpMV kernel is implemented by CUDA code. The steps to reproduce AlphaSparse is following:

\begin{enumerate} 
\item Clone the GitHub repository. And go to the root directory of AlphaSparse.
\item Set the location of nvcc used by AlphaSparse in ``cuda\_code/make\_template.sh''. AlphaSparse will use this shell script to compile the kernel generated by it.
\item Set the configuration of AlphaSparse. Fields named ``ROOT\_PATH\_STR'' and ``spmv\_header\_file'' in the file named ``global\_config.json'' needs set according to the path of AlphaSparse.
\item Create a directory for temporary data: ``mkdir data\_source''
\item Compile. ``make -j 16''
\item Download Matrix Market File from SuiteSparse Matrix Collection (https://sparse.tamu.edu).
\item Unpack the file. And use ``data\_prepare.py'' to preprocess it: ``python3 data\_prepare.py \{directory\_matrix\_market\_file\}/\{matrix\_name\}.mtx \{directory\_matrix\_market\_file\}/\{matrix\_name\}.mtx.coo''.
\item Design SpMV program for specific matrix: ``./main {directory\_matrix\_market\_file}/{matrix\_name}.mtx.coo > data\_source/test.log''
\item After several hours. The description of Operator Graph and its performance (GFLOPS) are shown in data\_source/test.log.  And corresponding SpMV kernel is shown in ``cuda\_code/template.cu''.
\end{enumerate}

We have also provided a Python script named ``batch\_test\_spmv\_builder.py'' for batch tests. A more detailed tutorial is shown in https://github.com/AnonymousRepo123/AlphaSparse.

\section*{System environment}

Except for the GPU, the environments of RTX 2080 and A100 platforms are the same. Our evaluation environment follows.

Linux: 5.4.0-99-generic Ubuntu 20.04 focal

CPU: Intel(R) Xeon(R) CPU E5-2620 v4 @ 2.10GHz

Memory: 64GB

Device: NVIDIA TU104 [GeForce RTX 2080]

Driver Version: 495.29.05    

CUDA Version: 11.5 

Python: 3.8

GCC: 9.4


\section*{Author-Created or Modified Artifacts:}

\begin{hangparas}{.25in}{1}

\textbf{Artifact 1}

Persistent ID: \ttvar{ https://github.com/AnonymousRepo123/AlphaSparse }

Artifact name: AlphaSparse
\end{hangparas}


\paragraph{Reproduction of the artifact without container:}
It is very easy to deploy AlphaSparse. No third-party library is needed to be installed by users. So it is not necessary to use a container.